\documentclass[12pt,titlepage]{article}

\begin{document}

\title{The peremptory influence of a uniform background for trapping neutral
fermions with an inversely linear potential}
\date{}
\author{Antonio S. de Castro \\
\\
Universidade de Coimbra\\
Centro de F\'{\i}sica Computacional\\
P-3004-516 Coimbra Portugal\\
and\\
UNESP - Campus de Guaratinguet\'{a}\\
Departamento de F\'{\i}sica e Qu\'{\i}mica\\
12516-410 Guaratinguet\'{a} SP - Brasil\\
\\
Electronic mail: castro@feg.unesp.br}
\maketitle

\begin{abstract}
The problem of neutral fermions subject to an inversely linear potential is
revisited. It is shown that an infinite set of bound-state solutions can be
found on the condition that the fermion is embedded in an additional uniform
background potential. An apparent paradox concerning the uncertainty
principle is solved by introducing the concept of effective Compton
wavelength.
\end{abstract}

\section{Introduction}

The four-dimensional Dirac equation with an anomalous magnetic-like (tensor)
coupling describes the interaction of neutral fermions with electric fields
and can be reduced to the two-dimensional Dirac equation with a pseudoscalar
coupling when the fermion is limited to move in just one direction.
Therefore, the investigation of the simpler Dirac equation in a 1+1
dimension with a pseudoscalar potential might be of relevance to a better
understanding of the problem of neutral fermions subject to electric fields
in the more realistic 3+1 world.

The bound states of fermions in one-plus-one dimensions by a pseudoscalar
double-step potential \cite{asc2} and their scattering by a pseudoscalar
step potential \cite{asc3} have already been analyzed in the literature
providing the opportunity to find some quite interesting results. Indeed,
the two-dimensional version of the anomalous magnetic-like interaction
linear in the radial coordinate, christened by Moshinsky and Szczepaniak
\cite{ms} as Dirac oscillator, has also received attention. Nogami and
Toyama \cite{nt}, Toyama et al. \cite{tplus} and Toyama and Nogami \cite{tn}
studied the behaviour of wave packets under the influence of that
conserving-parity potential whereas Szmytkowski and Gruchowski \cite{sg}
proved the completeness of the eigenfunctions. More recently Pacheco et al.
\cite{pa} studied some thermodynamics properties of the 1+1 dimensional
Dirac oscillator, and a generalization of the Dirac oscillator for a
negative coupling constant was presented in Ref. \cite{asc}. The
two-dimensional generalized Dirac oscillator plus an inversely linear
potential has also been addressed \cite{asc4}.

In recent papers, Villalba \cite{vil} and McKeon and Van Leeuwen \cite{mck}
considered a pseu\-do\-sca\-lar Coulomb potential ($V=\lambda /r$) in 3+1
dimensions and concluded that there are no bounded solutions. The reason
attributed in Ref. \cite{mck} for the absence of bounded solutions it that
the different parity eigenstates mix. Furthermore, the authors of Ref. \cite%
{mck} assert that \textit{the absence of bound states in this system
confuses the role of the }$\pi $\textit{-meson in the binding of nucleons}.
Such an intriguing conclusion sets the stage for the analyses by other sorts
of pseudoscalar potentials. A natural question to ask is if the absence of
bounded solutions by a pseudoscalar Coulomb potential is a characteristic
feature of the four-dimensional world. In Ref. \cite{asc} the Dirac equation
in one-plus-one dimensions with the pseudoscalar power-law potential $V=\mu
|x|^{\delta }$ was approached and there it was concluded that only for $%
\delta >0$ there can be a binding potential. That conclusion renders a sharp
contrast to the result found in \cite{mck} since Ref. \cite{asc} shows that
it is possible to find bound states for fermions interacting by a
pseudoscalar potential in 1+1 dimensions, to tell the truth there is
confinement, notwithstanding the spinor is not an eigenfunction of the
parity operator. One might ponder that the underlying reason is the way the
spinors are affected by the behaviour of the potentials at the origin as
well as at infinity because this is the difference between the Coulomb
potentials in those two dissimilar worlds. Nevertheless, two more recent
works show that it is not the case.

It was shown in Ref. \cite{asc10} that the presence of a uniform background
potential is a sine qua non condition for furnishing bounded solutions for
the pseudoscalar screened Coulomb potential ($\sim e^{-|x|/\lambda }$). This
last interesting work encourages the inclusion of a uniform background for
other sorts of potentials which otherwise are not able to hold bounded
solutions. The parity-violating inversely linear potential ($1/|x|$) is not
a binding potential \cite{asc} and the purpose of the present work to
investigate the influence of a uniform background on its spectrum. We show
that an infinite set of bounded solutions can come into existence because
the fermion embedded in the uniform background acquires both effective mass
and effective coupling constant. Beyond its importance as a new solution for
a fundamental equation in physics, the problem analyzed in the present work
adds a new contrast to the conclusions in Ref. \cite{mck}. Furthermore, it
shows the decisive and masterful influence of a uniform background to
furnish an infinite set of bound-state solutions.

\section{The Dirac equation with a pseudoscalar potential in a 1+1 dimension}

The 1+1 dimensional time-independent Dirac equation for a fermion of rest
mass $m$ coupled to a pseudoscalar potential reads

\begin{equation}
H\psi =E\psi ,\quad H=c\alpha p+\beta mc^{2}+\beta \gamma ^{5}V  \label{1}
\end{equation}

\noindent where $E$ is the energy of the fermion, $c$ is the velocity of
light and $p$ is the momentum operator. We use $\alpha =\sigma _{1}$ and $%
\beta =\sigma _{3}$, where $\sigma _{1}$ and $\sigma _{3}$ are Pauli
matrices, and $\beta \gamma ^{5}=\sigma _{2}$. Provided that the spinor is
written in terms of the upper and the lower components, $\psi _{+}$ and $%
\psi _{-}$ respectively, \noindent the Dirac equation decomposes into:

\begin{equation}
\left( -E\pm mc^{2}\right) \psi _{\pm }=i\hbar c\psi _{\mp }^{\prime }\pm
iV\psi _{\mp }  \label{2}
\end{equation}

\noindent where the prime denotes differentiation with respect to $x$. In
terms of $\psi _{+}$ and $\psi _{-}$ the spinor is normalized as $%
\int_{-\infty }^{+\infty }dx\left( |\psi _{+}|^{2}+|\psi _{-}|^{2}\right) =1$
so that $\psi _{+}$ and $\psi _{-}$ are square integrable functions.

In the nonrelativistic approximation (potential energies small compared to $%
mc^{2}$ and $E\approx mc^{2}$) Eq. (\ref{1}) becomes

\begin{equation}
\psi _{-}=\left( \frac{p}{2mc}\,+i\,\frac{V}{2mc^{2}}\right) \psi _{+}
\label{3}
\end{equation}

\begin{equation}
\left( -\frac{\hbar ^{2}}{2m}\frac{d^{2}}{dx^{2}}+\frac{V^{2}}{2mc^{2}}+%
\frac{\hbar V^{\prime }}{2mc}\right) \psi _{+}=\left( E-mc^{2}\right) \psi
_{+}  \label{4}
\end{equation}

\noindent Eq. (\ref{3}) shows that $\psi _{-}$ is of order $v/c<<1$ relative
to $\psi _{+}$ and Eq. (\ref{4}) shows that $\psi _{+}$ obeys the Schr\"{o}%
dinger equation (at this point the author digress to make his apologies for
mentioning in former papers (\cite{asc2}-\cite{asc3}) that the pseudoscalar
potential does not present any contributions in the nonrelativistic limit).
Note that the Dirac equation is not invariant under $V\rightarrow V+$const.
Therefore, the absolute values of the energy have physical significance and
the freedom to choose a zero-energy is lost. This last statement remains
truthfully in the nonrelativistic limit. It is also noticeable that the
pseudoscalar coupling results in the Schr\"{o}dinger equation with an
effective potential in the nonrelativistic limit, and not with the original
potential itself. Indeed, this is the side effect which in a 3+1 dimensional
space-time makes the linear potential to manifest itself as a harmonic
oscillator plus a strong spin-orbit coupling in the nonrelativistic limit
\cite{ms}. The form in which the original potential appears in the effective
potential, the $V^{2}$ term, allows us to infer that even a potential
unbounded from below could be a confining potential. This phenomenon is
inconceivable if one starts with the original potential in the
nonrelativistic equation. It has already been verified that a constant added
to the screened Coulomb potential is undoubtedly physically relevant \cite%
{asc10}. As a matter of fact, it plays a crucial role to ensure the
existence of bounded solutions. Nevertheless, the resulting potential does
not present any nonrelativistic limit.

For $E\neq \pm mc^{2}$, the coupling between the upper and the lower
components of the Dirac spinor can be formally eliminated when Eq. (\ref{2})
is written as second-order differential equations:

\begin{equation}
-\frac{\hbar ^{2}}{2}\;\psi _{\pm }^{\prime \prime }+\left( \frac{V^{2}}{%
2c^{2}}\pm \frac{\hbar }{2c}V^{\prime }\right) \;\psi _{\pm }=\frac{%
E^{2}-m^{2}c^{4}}{2c^{2}}\;\psi _{\pm }  \label{5}
\end{equation}

\noindent These last results show that the solution for this class of
problem consists in searching for bounded solutions for two Schr\"{o}dinger
equations. It should not be forgotten, though, that the equations for $\psi
_{+}$ or $\psi _{-}$ are not indeed independent because $E$ appears in both
equations. Therefore, one has to search for bound-state solutions for both
signals in (\ref{5}) with a common eigenvalue. At this stage on can realize
that \noindent \noindent the Dirac energy levels are symmetrical about $E=0$%
. It means that the potential couples to the positive-energy component of
the spinor in the same way it couples to the negative-energy component. In
other words, this sort of potential couples to the mass of the fermion
instead of its charge so that there is no atmosphere for the spontaneous
production of particle-antiparticle pairs. No matter the intensity and sign
of the potential, the positive- and the negative-energy solutions never
meet. Thus there is no room for transitions from positive- to
negative-energy solutions. This all means that Klein\'{}s paradox never
comes to the scenario. The solutions for $E=\pm mc^{2}$, excluded from the
Sturm-Liouville problem, can be obtained directly from the Dirac equation (%
\ref{2}).

The solutions for $E=\pm mc^{2}$, excluded from the Sturm-Liouville problem,
can be obtained directly from the Dirac equation (\ref{2}). One can observe
that such sort of isolated solutions can be written for $E=+mc^{2}$ as

\begin{eqnarray}
\psi _{-} &=&N_{-}\,\exp \left[ -v(x)\right]  \nonumber \\
&&  \label{6} \\
\psi _{+}^{\prime }-v^{\prime }\psi _{+} &=&+i\,\frac{2mc}{\hbar }%
N_{-}\,\exp \left[ -v(x)\right]  \nonumber
\end{eqnarray}

\noindent and for $E=-mc^{2}$ as

\begin{eqnarray}
\psi _{+} &=&N_{+}\,\exp \left[ +v(x)\right]  \nonumber \\
&&  \label{6a} \\
\psi _{-}^{\prime }+v^{\prime }\psi _{-} &=&-i\,\frac{2mc}{\hbar }%
N_{+}\,\exp \left[ +v(x)\right]  \nonumber
\end{eqnarray}

\noindent where $N_{+}$ and $N_{-}$ are normalization constants and $%
v(x)=\int^{x}dy\,V(y)\,/(\hbar c)$. \noindent Of course well-behaved
eigenstates are possible only if $v(x)$ has a distinctive leading asymptotic
behaviour.

\section{The inversely linear potential plus a uniform background}

Now let us concentrate our attention on the potential in the form
\begin{equation}
V=-\frac{\hbar cq}{|x|}+V_{0}  \label{7}
\end{equation}
\noindent where $V_{0}$ and the dimensionless coupling constant, $q$, are
real numbers. In this case the possible isolated solutions corresponding to $%
E=\pm mc^{2}$ are excluded from our consideration because they do not
fulfill the conditions of continuity and normalizability simultaneously. On
the other side, the Sturm-Liouville problem corresponding to Eq. (\ref{5})
becomes

\begin{equation}
H_{eff}\psi _{\pm }=-\frac{\hbar ^{2}}{2m_{eff}}\,\psi _{\pm }^{\prime
\prime }+V_{eff}^{[\pm ]}\,\psi _{\pm }=E_{eff}\,\psi _{\pm }  \label{8}
\end{equation}

\noindent where

\begin{equation}
V_{eff}^{[\pm ]}(x)=-\frac{\hbar cq_{eff}}{|x|}+\frac{A_{\pm }(x)}{x^{2}}%
,\quad A_{\pm }(x)=\frac{\hbar ^{2}}{2m_{eff}}\,q\left[ q\mp \mathrm{sgn}(x)%
\right]  \label{9}
\end{equation}

\noindent and

\begin{equation}
E_{eff}=\frac{E^{2}-m_{eff}^{2}c^{4}}{2m_{eff}c^{2}},\quad m_{eff}=m\sqrt{1+%
\frac{V_{0}^{2}}{m^{2}c^{4}}},\quad q_{eff}=q\,\frac{V_{0}}{m_{eff}c^{2}}
\label{10}
\end{equation}

\noindent Therefore, one has to search for bounded solutions of a particle
in an effective Kratzer-like potential \cite{kra}.

\subsection{Qualitative analysis}

Before proceeding, it is useful to make some qualitative arguments regarding
the Kratzer-like potential and its possible solutions. Although the
potential given by (\ref{7}) with $q<0$ gives rise to an ubiquitous
repulsive potential in a nonrelativistic theory, the possibility of such a
sort of potential to bind fermions, if $V_{0}<0$, is already noticeable in
the nonrelativistic limit of the Dirac equation (see Eq. (\ref{4})).
Furthermore, $V_{0}$ must be different from zero, otherwise there would be a
repulsive effective potential as long as the condition $A_{+}<0$ and $%
A_{-}<0 $ is never satisfied simultaneously. The parameters of the effective
potential with $V_{0}\neq 0$ and $q_{eff}>0$ fullfil the key conditions to
furnish spectra discrete with $E_{eff}<0$, corresponding to $|E|<mc^{2}$.
The Dirac eigenenergies belonging to $|E|>mc^{2}$ correspond to the
continuum. Note that the parameters of the effective potential are related
in such a manner that the change $q\rightarrow -q$ induces the change $%
A_{\pm }\rightarrow A_{\mp }$. The combined transformation $q\rightarrow -q$
and $V_{0}\rightarrow -V_{0}$ has as effect $V_{eff}^{[\pm ]}\rightarrow
V_{eff}^{[\mp ]}$, meaning that the effective potential for $\psi _{+}$
transforms into that one for $\psi _{-}$ and vice versa. On the other hand,
the change $x\rightarrow -x$ induces the change $V_{eff}^{[\pm
]}(-x)\rightarrow V_{eff}^{[\mp ]}(x)$ ($A_{\pm }(-x)\rightarrow A_{\mp }(x)$%
), implying that $|\psi _{\pm }(-x)|$ behaves like $|\psi _{\mp }(x)|$. One
can see that $\psi _{\pm }$ is subject to a potential-well structure for $%
V_{eff}^{[\pm ]}$ when $|q|>1$. For $|q|\leq 1$ the effective potential has
a potential-well structure on one side of the $x$-axis and a singular at the
origin on the other side. The singularity is given by $-1/|x|$ when $|q|=1$,
and $-1/|x|^{2}$ when $|q|<1$. It is worthwhile to mention at this point
that the singularity at $x=0$ never menaces the fermion to collapse to the
center \cite{lan} because in any condition $A_{\pm }$ is never less than the
critical value $A_{c}=-\hbar ^{2}/(8m_{eff})$. The Schr\"{o}dinger equation
with the Kratzer-like potential is an exactly solvable problem and its
solution, for a repulsive inverse-square term in the potential ($A_{\pm }>0$%
), can be found on textbooks \cite{lan}-\cite{flu}. Since we need solutions
involving a repulsive as well as an attractive inverse-square term in the
potential, the calculation including this generalization is presented.

Since $|\psi _{\pm }(-x)|$ behaves like $|\psi _{\mp }(x)|$ we can
concentrate our attention on the half-line and impose boundary conditions on
$\psi _{\pm }$ at $x=0$ and $x=\infty $. Square-integrability requires that $%
\psi _{\pm }(\infty )=0$ and the boundary condition at the origin comes into
existence by demanding that the effective Hamiltonian given (\ref{8}) is
Hermitian, viz.

\begin{equation}
\int_{0}^{\infty }dx\;\psi _{k}^{*}\left( H_{eff}\psi _{k^{^{\prime
}}}\right) =\int_{0}^{\infty }dx\;\left( H_{eff}\psi _{k}\right) ^{*}\psi
_{k^{^{\prime }}}  \label{15}
\end{equation}

\noindent where $\psi _{k}$ is an eigenfunction corresponding to an
effective eigenvalue $\left( E_{eff}\right) _{k}$. In passing, note that a
necessary consequence of Eq. (\ref{15}) is that the eigenfunctions
corresponding to distinct effective eigenvalues are orthogonal. It can be
shown that (\ref{15}) is equivalent to

\begin{equation}
\lim_{x\rightarrow 0}\left( \psi _{k}^{*}\frac{d\psi _{k^{^{\prime }}}}{dx}-%
\frac{d\psi _{k}^{*}}{dx}\psi _{k^{^{\prime }}}\right) =0  \label{16}
\end{equation}

\subsection{Quantitative analysis}

According to the previous qualitative analysis, it is convenient to define
the dimensionless quantities $z$ and $B$,

\negthinspace
\begin{equation}
z=\frac{2}{\hbar }\sqrt{-2m_{eff}E_{eff}}\;|x|,\quad B=q_{eff}\,\sqrt{-\frac{%
m_{eff}c^{2}}{2E_{eff}}}  \label{11}
\end{equation}

\noindent and reduce (\ref{8}) to the form

\begin{equation}
\,\psi _{\pm }^{\prime \prime }+\left( -\frac{1}{4}+\frac{B}{z}-\frac{%
2m_{eff}A_{\pm }}{\hbar ^{2}z^{2}}\right) \psi _{\pm }=0  \label{12}
\end{equation}

\noindent Now the prime denotes differentiation with respect to $z$. The
normalizable asymptotic form of the solution as $z\rightarrow \infty $ is $%
e^{-z/2}$. As $z\rightarrow 0$, when the term $1/z^{2}$ dominates, the
solution behaves as $z^{s_{\pm }}$, where $s_{\pm }$ is a solution of the
algebraic equation

\begin{equation}
s_{\pm }(s_{\pm }-1)-\frac{2m_{eff}A_{\pm }}{\hbar ^{2}}=0  \label{13}
\end{equation}
The boundary condition at $x=0$ demands $s_{\pm }\geq 1/2$ so that the
solution of (\ref{13}) is given by

\begin{equation}
s_{\pm }=\frac{1}{2}\left( 1+\sqrt{1+\frac{8m_{eff}A_{\pm }}{\hbar ^{2}}}%
\right) =\frac{1}{2}+|q\mp \frac{1}{2}|  \label{14}
\end{equation}
The solution for all $z$ can be expressed as $\psi _{\pm }=z^{s_{\pm
}}e^{-z/2}w_{\pm }$, where $w_{\pm }$ is solution of Kummer\'{}s equation
\cite{abr}

\begin{equation}
zw_{\pm }^{\prime \prime }+(b_{\pm }-z)w_{\pm }^{\prime }-a_{\pm }w_{\pm }=0
\label{17}
\end{equation}

\noindent with

\begin{equation}
a_{\pm }=s_{\pm }-B,\quad b_{\pm }=2s_{\pm }  \label{18}
\end{equation}

\noindent Then $w_{\pm }$ is expressed as $M(a_{\pm },b_{\pm },z)$ and in
order to furnish normalizable $\psi _{\pm }$, the confluent hypergeometric
function must be a polynomial. This demands that $a_{\pm }=-n_{\pm }$, where
$n_{\pm }$ is a nonnegative integer in such a way that $B>0$ (corresponding
to $q_{eff}>0$) and $M(a_{\pm },b_{\pm },z)$ is proportional to the
associated Laguerre polynomial $L_{n_{\pm }}^{b_{\pm }-1}(z)$, a polynomial
of degree $n_{\pm }$. This requirement, combined with the first equation of (%
\ref{18}), also implies into quantized effective eigenvalues:

\begin{equation}
E_{eff}=-\,m_{eff}c^{2}\,\frac{q_{eff}^{2}}{2\left( s_{\pm }+n_{\pm }\right)
^{2}},\qquad n_{\pm }=0,1,2,\ldots  \label{21}
\end{equation}

\noindent with eigenfunctions given by

\begin{equation}
\psi _{\pm }=N_{_{\pm }}\;z^{s_{\pm }}\,e_{\;}^{-z/2}\;L_{n_{\pm }}^{2s_{\pm
}-1}\left( z\right)  \label{22}
\end{equation}

\smallskip

\noindent

\noindent where $N_{_{\pm }}$ is a normalization constant. Note that the
behaviour of $\psi _{\pm }$ at very small $z$ implies into the Dirichlet
boundary condition $\psi _{\pm }(0)=0$. This boundary condition is essential
whenever $A_{\pm }\neq 0$, nevertheless it also develops for $A_{\pm }=0$.

The necessary conditions for binding fermions in the Dirac equation with the
effective Kratzer-like potential have been put forward. The formal
analytical solutions have also been obtained. Now we move on to consider a
survey for distinct cases in order to match the common effective eigenvalue.
As we will see this survey leads to additional restrictions on the
solutions, including constraints involving the nodal structure of the Dirac
spinor.

From (\ref{14}) one sees that for $q\geq 1/2$ one has $s_{+}=q$ and $%
s_{-}=s_{+}+1$, for $q\leq -1/2$ one has $s_{+}=-q+1$ and $s_{-}=s_{+}-1$,
and for $-1/2<q<+1/2$ one has $s_{+}=-q+1$ and $s_{-}=s_{+}+2$. Therefore,
demanding a common eigenvalue implies that for $q\geq 1/2$ ($q\leq -1/2$)
one has $n_{-}=n-1$ ($n_{-}=n+1$), where $n=n_{+}$. On the other hand, for $%
-1/2<q<+1/2$ one has $n_{-}=n-2q$, showing to be an unacceptable possibility
because it does not provide an integer value for $n-n_{-}$. As an immediate
consequence of this analysis one can see that the solutions split into two
distinct classes. In order to write $\psi _{\pm }$ on the whole line we
recur again to the observation that $|\psi _{\pm }(-x)|$ behaves like $|\psi
_{\mp }(x)|$. Nevertheless, the matter is a little more complicated because
the potential presents a singularity at the origin so that $\psi _{\pm }$
can present a discontinuity there. \noindent In fact, the first-order
differential equation given by (\ref{2}) implies that $\psi _{\pm }$ can be
discontinuous wherever the potential undergoes an infinite jump. In the
specific case under consideration, the effect of the singularity of the
potential can be evaluated by integrating (\ref{2}) from $-\delta $ to $%
+\delta $ and taking the limit $\delta \rightarrow 0$. The connection
condition relating $\psi _{\pm }(+\delta )$ and $\psi _{\pm }(-\delta )$ can
be summarized as

\begin{equation}
\psi _{\pm }(+\delta )-\psi _{\pm }(-\delta )=\mp q\int_{-\delta }^{+\delta
}dx\;\frac{\psi _{\pm }}{|x|}  \label{23f}
\end{equation}
Substitution of (\ref{22}) into (\ref{23f}) allows us to conclude that $\psi
_{\pm }(+\delta )=\psi _{\pm }(-\delta )$ in all the circumstances. The
continuity of the spinor at the origin does $\psi _{\pm }(-x)$ to differ
from $\psi _{\mp }(x)$ just by a factor related to the relative
normalization. Therefore,

\bigskip

\textbf{A)} For $q\geq 1/2$ ($V_{0}>0$):

\[
E=\pm mc^{2}\,\sqrt{1+\frac{V_{0}^{2}}{m^{2}c^{4}}\left[ 1-\left( \frac{q}{%
n+q}\right) ^{2}\right] },\qquad n=1,2,3,\ldots
\]

\begin{equation}
\psi _{+}=\,e_{\;}^{-z/2}\left[ \theta (-x)\;z^{q+1}\;L_{n-1}^{2q+1}\left(
z\right) +\theta (+x)\;z^{q}\;L_{n}^{2q-1}\left( z\right) \right]  \label{23}
\end{equation}

\[
\psi _{-}=N\,e_{\;}^{-z/2}\left[ \theta (-x)\;z^{q}\;L_{n}^{2q-1}\left(
z\right) +\theta (+x)\;z^{q+1}\;L_{n-1}^{2q+1}\left( z\right) \right]
\]

\bigskip

\textbf{B)} For $q\leq -1/2$ ($V_{0}<0$):
\[
E=\pm mc^{2}\,\sqrt{1+\frac{V_{0}^{2}}{m^{2}c^{4}}\left[ 1-\left( \frac{q}{%
n-q+1}\right) ^{2}\right] },\qquad n=0,1,2,\ldots
\]

\begin{equation}
\psi _{+}=\,e_{\;}^{-z/2}\left[ \theta (-x)\;z^{-q}\;L_{n+1}^{-2q-1}\left(
z\right) +\theta (+x)\;z^{-q+1}\;L_{n}^{-2q+1}\left( z\right) \right]
\label{24}
\end{equation}

\[
\psi _{-}=N\,\,e_{\;}^{-z/2}\left[ \theta
(-x)\;z^{-q+1}\;L_{n}^{-2q+1}\left( z\right) +\theta
(+x)\;z^{-q}\;L_{n+1}^{-2q-1}\left( z\right) \right]
\]

\bigskip

\noindent where $N$ is a normalization constant and $\theta (x)$ is the
Heaviside step function. The preceding results show explicitly that for any $%
V_{0}\neq 0$ (recall that $\mathrm{sgn}\left( qV_{0}\right) >0$) there is an
infinite set of bound-state solutions with eigenenergies into the spectral
gap between $-mc^{2}$ and $+mc^{2}$ arranged symmetrically about $E=0$. It
is also evident that the ultrarelativistic zero-eigenmodes are allowed as $%
|q|\rightarrow \infty $. If $V_{0}=0$, though, there are no bounded
solutions at all, as predicted by the qualitative arguments.

\section{Conclusions}

We have succeed in searching for exact bounded solutions for massive neutral
fermions by considering a violating-parity inversely linear potential plus a
uniform background, despite the mixing of parities. A remarkable feature of
this problem is the possibility of trapping a neutral fermion with an
uncertainty in the position that can shrink without limit as $|V_{0}|$
increases. At first glance it seems that the uncertainty principle dies away
provided such a principle implies that it is impossible to localize a
particle into a region of space less than half of its Compton wavelength
(see, e.g., Ref. \cite{str}). However, a result consistent with the
uncertainty principle can be obtained by using the effective Compton
wavelength $\lambda _{c}=\hbar /(m_{eff}c$). It means that the localization
of a fermion under the influence of this sort of pseudoscalar potential does
not require any minimum value in order to ensure the single-particle
interpretation of the Dirac equation.

\bigskip \bigskip

\noindent \textbf{Acknowledgments}

This work was supported in part by means of funds provided by CNPq and
FAPESP.

\end{document}